\documentclass[aip,jcp,reprint]{revtex4-1}

\usepackage{dcolumn}       
\usepackage{bm}            
\usepackage{amsmath}       
\usepackage{graphicx}      
\usepackage{natbib}                
\usepackage{tabularx}
\usepackage{amssymb}       
\usepackage{latexsym}      
\usepackage{multirow}      
\usepackage{array}         
\usepackage{wrapfig}       
\usepackage{textcomp}
\usepackage{color}

\newcommand{\Figref}[1]{Fig.~\ref{#1}}
\newcommand{\Eqref}[1]{Eq.~(\ref{#1})}
\newcommand{\ie}{\emph{i.e.}}
\newcommand{\eg}{\emph{e.g.}}
\newcommand{\inter}{\emph{inter}}
\newcommand{\intra}{\emph{intra}}
\newcommand{\Inter}{\emph{Inter}}
\newcommand{\Intra}{\emph{Intra}}

\begin{document}

\title{A tunable electronic beam splitter realized with crossed graphene nanoribbons}

\author{Pedro Brandimarte}
\email{brandimarte@gmail.com}
\affiliation{Centro de F\'{\i}sica de Materiales (CFM) CSIC-UPV/EHU,  Paseo Manuel de Lardizabal 5, E-20018, Donostia-San~Sebasti\'an, Spain}

\author{Mads~Engelund}
\affiliation{Centro de F\'{\i}sica de Materiales (CFM) CSIC-UPV/EHU,  Paseo Manuel de Lardizabal 5, E-20018, Donostia-San~Sebasti\'an, Spain}

\author{Nick~Papior}
\affiliation{Institut de Ci\`encia de Materials de Barcelona (ICMAB-CSIC), Campus de la UAB, E-08193 Bellaterra, Spain}

\author{Aran~Garcia-Lekue}
\affiliation{Donostia International Physics Center, DIPC, Paseo Manuel de Lardizabal 4, E-20018, 
Donostia-San~Sebasti\'an, Spain}
\affiliation{IKERBASQUE, Basque Foundation for Science, E-48013, Bilbao, Spain}

\author{Thomas~Frederiksen}
\email[Corresponding author: ]{thomas\_frederiksen@ehu.eus}
\affiliation{Donostia International Physics Center, DIPC, Paseo Manuel de Lardizabal 4, E-20018, 
Donostia-San~Sebasti\'an, Spain}
\affiliation{IKERBASQUE, Basque Foundation for Science, E-48013, Bilbao, Spain}

\author{Daniel~S\'anchez-Portal}
\affiliation{Centro de F\'{\i}sica de Materiales (CFM) CSIC-UPV/EHU,  Paseo Manuel de Lardizabal 5, E-20018, Donostia-San~Sebasti\'an, Spain}
\affiliation{Donostia International Physics Center, DIPC, Paseo Manuel de Lardizabal 4, E-20018,
Donostia-San~Sebasti\'an, Spain}

\date{\today}

\begin{abstract}
Graphene nanoribbons (GNRs) are promising components in future nanoelectronics due to the large mobility of graphene electrons and their tunable electronic band gap in combination with recent experimental developments of on-surface chemistry strategies for their growth. 
Here we explore a prototype 4-terminal semiconducting device formed by two crossed armchair GNRs (AGNRs) using state-of-the-art 
\emph{first-principles} 
transport methods.
We analyze in detail the roles of intersection angle, stacking order, \inter-GNR separation, and finite voltages on the transport characteristics.
Interestingly, when the AGNRs intersect at $\theta=60^\circ$, electrons injected from one terminal can be split into two outgoing waves with a tunable ratio around 50\% and with almost negligible back-reflection.
The splitted electron wave is found to propagate partly straight across the intersection region in one ribbon and partly in \emph{one} direction of the other ribbon, \ie, in analogy of an optical beam splitter. 
Our simulations further identify realistic conditions for which this semiconducting device can act as a mechanically controllable electronic beam splitter with possible applications in carbon-based quantum electronic circuits and electron optics.
We rationalize our findings with a simple model that suggests
that electronic beam splitters can generally be realized with crossed GNRs.
\end{abstract}

\maketitle

\section{Introduction}

The wave nature of electrons that propagate coherently in ballistic, one-dimensional conductors has certain qualities in common with photons propagating in vacuum.\cite{Bocquillon2014} 
This analogy has spawned the field of electron quantum optics, in which a number of optical setups have been realized in form of their electronic counterparts, such as the Hanbury Brown and Twiss geometry for studies of Fermion anti-bunching \cite{Henny1999,Oliver1999} and the two-particle Aharanov-Bohm effect \cite{Samuelsson2004,Neder2007,Splettstoesser2010} as well as Mach--Zehnder interferometry with charged quasiparticles \cite{Ji2003,Roulleau2007}.
The advent of coherent single-particle sources \cite{Feve2007,Dubois2013,Bocquillon2013, Waldie2015, Ryu2016} and entangled electron pair generators\cite{Hofstetter2009, Ubbelohde2014}
has further provided exciting possibilities for novel quantum technologies and information processing.

A fundamental component for such electron quantum optics is the need for semi-transparent ``mirrors'', \ie, electronic beam splitters.
Currently, most experiments\cite{Bocquillon2014} rely on mesoscopic devices based on high-mobility two-dimensional electron gases in the quantum Hall effect regime, in which the electron transport occurs by chiral edge channels that are generally protected against backscattering. A beam splitter is here realized with a quantum point contact that is tuned via electrostatic gates such that only one quantum transport channel transmits with probability $T=0.5$. However, a drawback of the technology in quantum Hall regime is the need for low temperatures and high magnetic fields which severely limits possible applications outside of the laboratory.

Graphene nanoribbons (GNRs)~\cite{Fujita1996,Nakada1996,Wakabayashi1999} have some highly desirable properties for their use in molecular-scale electronics devices -- they can be designed with specific band gaps~\cite{Han2007,Son2006,Yang2007} and long defect-free samples can now be fabricated with both armchair (AGNR)~\cite{Cai2010} and zigzag (ZGNR) edge topology~\cite{Ruffieux2016} via on-surface synthesis.
However, in the standard bottom-up approach it is difficult to fully explore the GNR electronic properties due to interactions with the metallic substrates used for the synthesis. Very recently this drawback has been bypassed using synthesis on a semi-conducting substrate~\cite{Oliveira2015,Jacobberger2015} and by post-synthesis transfer to an insulating substrate~\cite{Ruffieux2016}.
Manipulation of single GNRs have also been demonstrated with scanning probe microscopy,~\cite{Koch2012,Kawai2016} which opens the possibility to built novel electronic networks with GNRs.
Simple 4-terminal tunneling junctions can be fabricated by crossing 1D-structures such as carbon nanotubes~\cite{Fuhrer2000,Yoon2001} or GNRs.~\cite{Jiao2010}
Indeed, in the context of electron quantum optics, it was very recently theoretically proposed that two crossed ZGNRs could act as an electronic beam splitter \cite{Lima2016}.

The quantum transport properties of GNR-based devices have been extensively studied with first-principles methods, for instance in the contexts of chemical functionalization~\cite{Lopez2009}, optical excitations~\cite{Osella2012,Villegas2014}, thermoelectrics~\cite{Tan2011,Saha2011,Sevincli2013}, local current-density patterns~\cite{Wilhelm2014}, vibrational excitations~\cite{Christensen2015}, and spin-scattering in ZGNRs~\cite{Kim2008,Zeng2010,Zeng2011} and hydrogenated AGNRs.~\cite{Soriano2010,Wilhelm2015}.
Various multi-terminal GNR geometries have also been addressed,
both in-plane GNR devices~\cite{Jayasekera2007, Areshkin2007, Botello2011, Xu2013} and tunneling junctions formed between  GNRs.~\cite{Botello2011,Habib2012,Habib2013,Saha2013,VanDePut2016} 
Finite-bias calculations in a multi-terminal context were pioneered by Saha \emph{et al.}~\cite{Saha2009} and are becoming increasingly accessible in first-principles transport codes, such as the post-processing tool \textsc{Gollum}~\cite{Ferrer2014} and the open-source, self-consistent methods of \textsc{TranSiesta}.~\cite{Brandbyge2002, Papior2016}

In this manuscript we employ state-of-the-art \textit{first-principles} methods to study the transport properties of tunneling junctions formed by two crossed AGNRs. Earlier studies have explored similar systems,\cite{Habib2012, VanDePut2016} but these did not account for the charge redistribution in the junction at finite bias.
We analyze in detail the roles of intersection angle, stacking order, \inter-GNR separation, and finite voltages in this effective 4-terminal device. Interestingly, we discovered that 
when the two AGNRs cross at an intersection angle $\theta=60^\circ$ a substantial current can be passed from one ribbon to the other and,
more specifically, that electrons injected from one terminal can be split into two outgoing waves with a tunable ratio around 50\% and with almost negligible back-reflection.
We quantify how this \inter-GNR tunneling mechanism depends on the precise atomic arrangement and demonstrate how this enables our device to be tuned and controlled to act as an electronic beam splitter. 
We further propose a simple model to understand qualitatively the critical role of the intersection angle, which points toward the possibility that electronic beam splitters can be realized with GNRs of different chiralities and widths. We therefore speculate that such GNR-based
beam splitters could find applications in electron quantum optics at the nanoscale.

\section{Methodology}

\subsection{Multiterminal DFT-NEGF}

The calculations presented here were performed using the \textsc{Siesta}/\textsc{TranSiesta} packages~\cite{Soler2002,Brandbyge2002} that are based on density functional theory (DFT) and
nonequilibrium Green's functions (NEGF), a combination that is referred to as DFT-NEGF.
The \textsc{TranSiesta} code which was recently generalized to deal with multi-terminal devices in complex geometries, \ie, to allow any number of electrodes pointing in arbitrary directions~\cite{Papior2016}. 
Following Saha \emph{et al}.~\cite{Saha2009}, our multi-terminal system is defined by an expanded scattering region that includes the connections to the electrodes and a central region which is chosen such that any two terminals only interact through it.
Each semi-infinite terminal $j$ is assumed to be in thermal equilibrium characterized by a chemical potential $\mu_j$. 
The transport properties at the steady state are obtained within the NEGF approach~\cite{Keldysh1965,Kadanoff1962} by the propagator through the scattering region $G^r$ which, at energy $E$, is given by:
\begin{equation}
  G^{r} = \Big[ \varepsilon S-H - \sum^{N}_{j = 1}{\Sigma^{r}_{_j}} \Big]^{-1} \,,
\end{equation}
with $\varepsilon = \lim_{\eta \rightarrow 0^+} E + i \eta$. Here $S$ and $H$ are the scattering region overlap and Hamiltonian matrices, respectively, and $\Sigma^{r}_j$ the $j$-th lead retarded self-energy that introduces the effect of connecting the $j$-th electrode to the central region. On the one hand, when a bias voltage is applied to an electrode it is assumed that its energy levels are rigidly shifted. Therefore, each electrode $j$ has a chemical potential defined by $\mu_j = E_F + \alpha_j e V$, where $E_F$ is the Fermi energy of the combined system in equilibrium, $V$ is the applied bias window (the maximum absolute potential difference between any two terminals) and $\alpha_j \in [-0.5,0.5]$ is a proportionality factor that defines the chemical potential of the $j$-th electrode in terms of $V$. The central region, on the other hand, will have the charge distribution modified due to the connection to biased electrodes, which is then determined self-consistently within the DFT-NEGF procedure~\cite{Papior2016}.

In a multi-terminal setup it is a non-trivial task to determine the electrostatic potential which solves the Poisson equation and fulfills the boundary conditions imposed by all electrodes. In our calculations, we use the \emph{box} approximation,\cite{Papior2016} which consists of reinforcing the potential difference between the electrodes at each self-consistent step. This is done by adding the chemical potential $\mu_j$ to the periodic solution of the Poisson equation at the region belonging to the $j$-th electrode, and with a redefinition of the common energy reference at each iteration step.
The box approximation, particularly when combined with semi-conducting low-dimensional electrodes as in the present case, can potentially create an abrupt behavior of the potential at the boundaries between the electrodes and the central region.
However, in the calculations presented here, only a modest charge accumulation occurs at the central-region/lead boundary. Even at the largest applied bias less than $\pm 0.02 e^-$/atom accumulate at each side of the boundary, producing a negligible scattering, as demonstrated by the fact that varying the locations of the central-region/lead boundaries had only a negligible effect on the results. 

Once the DFT-NEGF self-consistency is achieved, one can compute the transport properties. The current flowing out of a given electrode $j$ depends on the transmission probabilities $T_{jj'}(E)$ of electrons being scattered to any of the other electrodes $j'$. This is expressed in terms of the multi-terminal Landauer-B\"{u}ttiker formula \cite{Buttiker1986}:
\begin{equation}
I_{_j} = \frac{2e}{h} \sum_{j'\neq j} \int \limits_{\scriptscriptstyle{-} \infty}^{\scriptscriptstyle{+} \infty} d E \, \underbrace{\textrm{Tr} \left[ A_{_j} \Gamma_{_{j'}} \right]}_{T_{jj'}} \left( f_{_j} - f_{_{j'}} \right) \,,
\end{equation}
where
\begin{equation}
A_{_j} = G^r \Gamma_j {G^{r}}^{\dagger}
\end{equation}
is the $j$-th electrode spectral function,
\begin{equation}
\Gamma_j = i \left[ \Sigma^{r}_j  - \left( \Sigma^{r}_j \right)^{\dagger} \right]
\end{equation}
the level-width function, and $f_j = f{(E - \mu_j)}$ the Fermi-Dirac distribution. The factor 2 is due to the spin degeneracy in a spin-less treatment.

Finally, to analyze the electron transport properties of multi-terminal devices in real space, we calculate the so-called \emph{bond currents}~\cite{Todorov2002}, \ie, the amount of current flowing from atom $\alpha$ to $\beta$.  
For scattering states originating from the $j$-th electrode the bond current $\mathcal{I}_{j,\alpha \beta}$ is defined as:

\begin{align}
  \label{eq:bond-current}
  \mathcal{I}_{j,\alpha \beta} &= i \frac{2 e}{h} \int_{E_a}^{E_b} dE
  \left[ H_{\alpha \beta} A_{j,\beta \alpha}-H_{ \beta\alpha} A_{j,\alpha\beta}  \right]\\
  &=-\mathcal{I}_{j,\beta\alpha} \,, \nonumber
\end{align}
where $[E_a,E_b]$ characterizes the energy window of interest. A summation over orbitals belonging to atoms $\alpha$ and $\beta$ is implicit in \Eqref{eq:bond-current}.

\subsection{Details of calculations}

\begin{figure} 
\includegraphics[width=.48\textwidth]{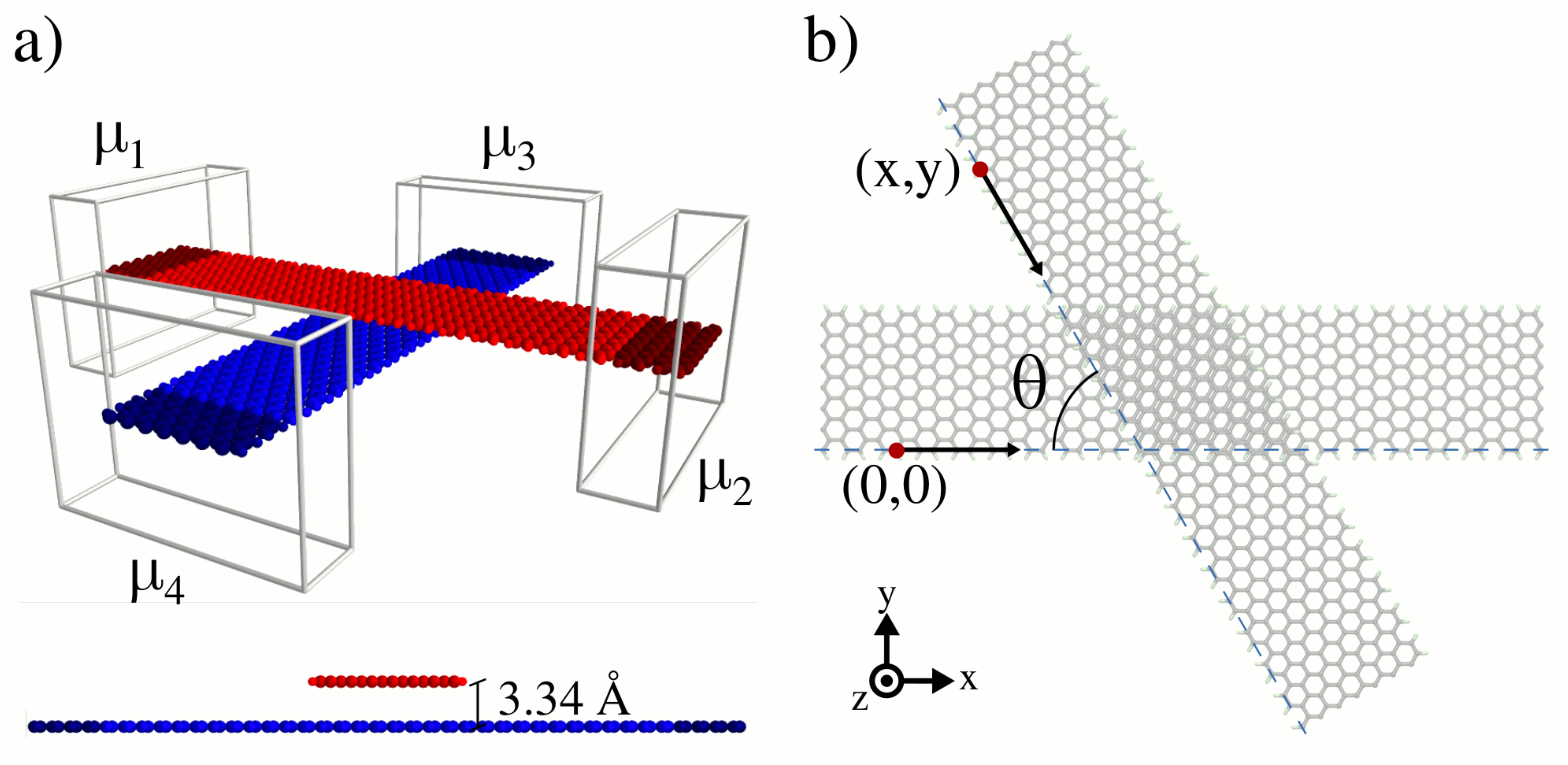}
\caption{Generic 4-terminal device setup formed by two crossed GNRs. (a) Two 14-AGNRs cross each other in a $\theta=90^\circ$ angle. The two planar GNRs are separated by a distance of $c=3.34$ {\AA} as shown in the side view. The system is coupled to four semi-infinite electrodes $j$, each with an independent chemical potential fixed at $\mu_j$. The boxed regions indicate those atoms which are considered as belonging to each biased electrode in the calculation. (b) Geometric parameters -- the angle $\theta$ and coordinates $(x,y)$ -- for an arbitrary crossbar geometry defined in terms of the edge vectors (black arrows) and the relative position of two edge C atoms (red dots).
}
\label{fig:system}
\end{figure}

Our device, shown in \Figref{fig:system}, is comprised of two infinite H-passivated 14-AGNRs (armchair graphene nanoribbons with a width of $W=14$ carbon atoms) rotated by an angle $\theta$ with respect to each other. 
Each GNR in the scattering region thus bridges two semi-infinite electrodes, one on each side of the intersection \ie, the system has a total of four terminals.
All calculations were therefore performed using the vdW density functional of Dion \emph{et al}~\cite{Dion2004} with the modified exchange by Klime\v{s}, Bowler and Michaelides~\cite{Klimes2010} since the description of dispersive interactions is crucial to describe interlayer binding and density rearrangment.~\cite{Santos2012}
The core electrons were described by non-local Troullier-Martins pseudopotentials~\cite{Troullier1991} and a double-$\zeta$ basis set was used to expand the valence-electron wavefunctions.~\cite{Soler2002} The fineness of the real space grid and the orbital radii were defined using, respectively, a 350 Ry energy cutoff and a 30 meV {\it energy shift}.~\cite{Soler2002}

First we allowed one ribbon, with axis along the $\hat{x}$ direction, to fully relax using conjugate gradient method using a force tolerance of 5 meV/\AA. The relaxed structure was then duplicated and translated along the $\hat{z}$ direction by 3.34~\AA\ (lowest energy distance for $\theta=90^\circ$) to explore the dependence of the transport properties on the other geometrical parameters (angle, stacking) defining our device.\footnote{Using the same functional and basis set, we found that for bilayer graphene the equilibrium distance $c$ between the two layers is 3.486~\AA\ for an AA stacking and 3.294~\AA\ for AB stacking. For a crossed GNR system with one GNR rotated by 90$^\circ$ with respect to the other, we found 3.339 \AA\ as the lowest energy distance, which lies in between these two extremes values for a bilayer graphene. Moreover, the lattice parameter calculated for graphite with different stackings ($c(\text{AAA})=3.469$ \AA, $c(\text{ABA})=3.348$ \AA\ and $c(\text{ABC})=3.310$ \AA), which are in close agreement with experiment measurements of $c=3.354$ \AA~\cite{Baskin1955,Zhao1989}, also presents an interlayer distance between the bilayer graphene values for AA and AB stacking. Therefore, we found reasonable to use the distance $c=3.339$ \AA\ as the reference value for all our crossed structures.}
We additionally considered the dependence of the transport properties on small variations of the distance between the ribbons. 

Each GNR consists of 640 atoms in the scattering region  and, altogether, the system is described by total of 9280 orbitals with the chosen basis set.
The electrode region $j$, \ie, where the $j$-th semi-infinite lead is coupled to the system (boxes in \Figref{fig:system}a), is defined by 64 atoms and is described by a chemical potential $\mu_j$.
The system configuration (relative position and rotation) is defined by the angle $\theta$ between the edge vectors (black arrows at \Figref{fig:system}b) and the relative position between one reference atom and its replica.
In order to uniquely define the different structures, we choose the reference atom as the fifth carbon atom along one edge (red dots at \Figref{fig:system}b).
The system is thus geometrically defined by $(x,y,\theta)$. In what follows, if only $\theta$ is explicitly specified then it is understood that the duplicated ribbon was rotated with respect the center of mass of the portion of the ribbon in the central region, \ie, that shown in \Figref{fig:system}.

In our simulations we considered an electronic temperature of $T=300$ K. For the electrode calculations we used 60 $k$-points along the periodic direction.
A level broadening of $\eta=10^{-7}$ eV was considered in the electrodes, while $\eta=10^{-5}$ eV was used for the contour integrations over the complex plane.~\cite{Papior2016,Brandbyge2002}
The self-consistency cycle was stopped when the difference between each element of the density matrix changed by less than $10^{-6}$.

\section{Results}

Throughout the paper we will use \intra-GNR to refer to events on the same ribbon (such as the transmission between electrodes belonging to the same ribbon) and \inter-GNR for events involving the two different ribbons (such as the transmission from one ribbon to the other).
Also, we will refer to the 14-AGNR attached to the electrodes 1 and 2 as GNR$_{12}$ and, analogously, the ribbon attached to the electrodes 3 and 4 as GNR$_{34}$.

\subsection{Band structure and zero-bias transmission of the isolated 14-AGNR}

A natural starting point for investigating the crossbar system lies in understanding the properties of a single GNR. As seen in \Figref{fig:trans90}a, our periodic calculations predict the 14-AGNR to be a semiconductor with a small band gap of $E_g=132$ meV, consistent with the result expected for a width of type $W=3p+2$.\cite{Son2006} This opening of the gap as compared to bulk graphene occurs due to the one-dimensional confinement. A direct correspondence between the band structure and the zero-bias transmission of a pristine 14-AGNR (\Figref{fig:trans90}b, dashed green line) can be made. As for any one-dimensional pristine structure of atomic-scale cross-section, the GNRs have a step-like transmission $T(E)=N_T(E)$, where $N_T(E)$ is the number of conductance channels (or, equivalently, number of bands) available at a given energy $E$. Thus, for the single 14-AGNR around the Fermi level (in a range of approximately $\pm 0.8$~eV) we first find a small region of zero transmission, due to the gap, and a plateau of transmission 1 associated with the highest valence (VB) and lowest conduction band (CB), respectively at larger negative and positive energies. The behavior of the transmission is very symmetric with respect to middle of the band gap, reflecting the approximate electron-hole symmetry of the band structure in the system.

\subsection{Effect of \inter-ribbon voltage}

\begin{figure*} 
\includegraphics[width=\textwidth]{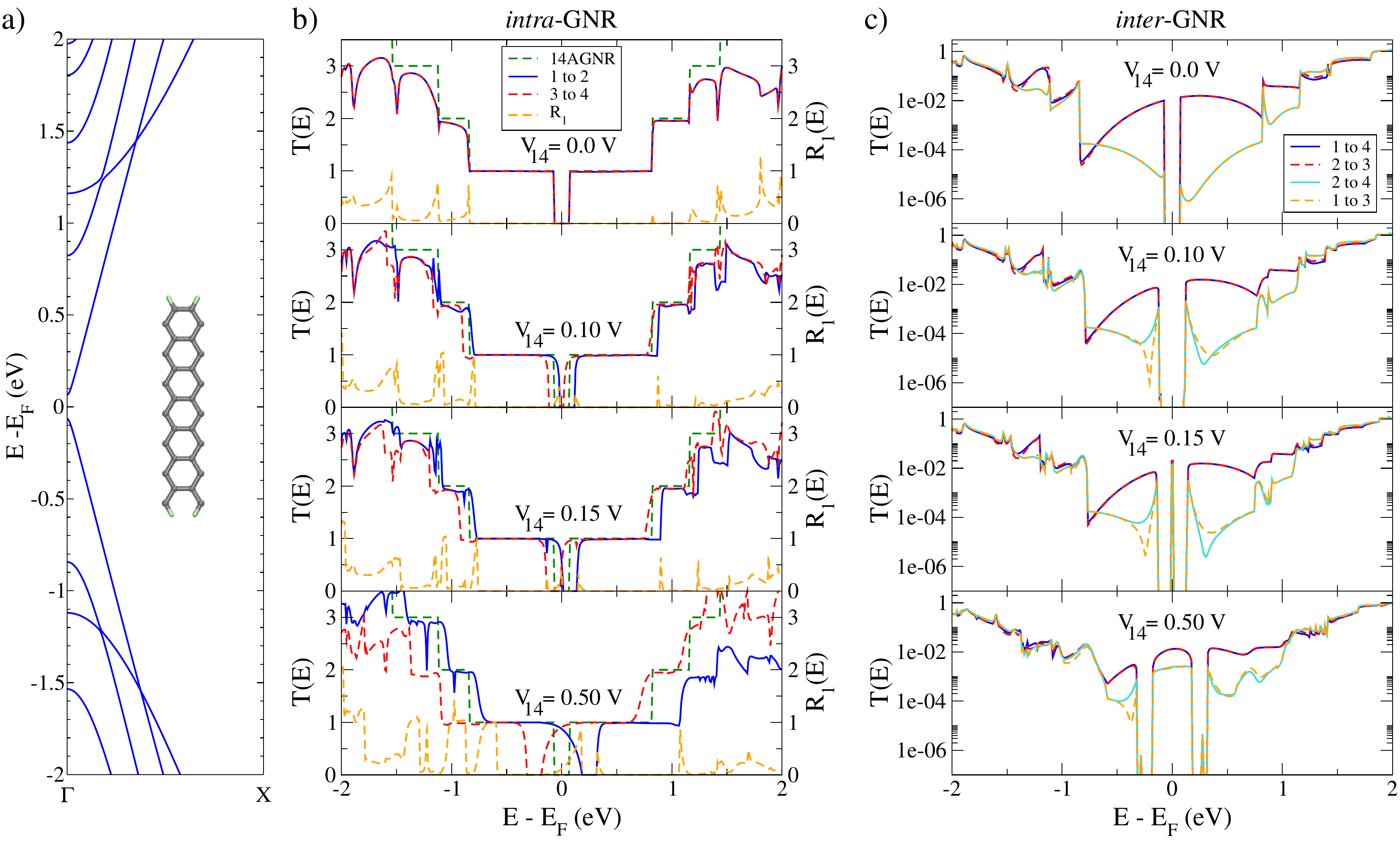}
\caption{Nonequilibrium transport characteristics for two AGNRs intersecting at $\theta=90^\circ$ $(x=25.72\text{\AA},y=41.79\text{\AA})$.
(a) Band structure of the bulk 14-AGNR electrodes (inset presents the corresponding unit cell). (b) \Intra-GNR transmissions $T_{12}$ and $T_{34}$ and reflection $R_1$ as a function of an \inter-ribbon voltage $V_{14}$ ($eV_{14}\equiv\mu_1-\mu_4$, with $\mu_1=\mu_2$ and $\mu_3=\mu_4$). Green dashed lines show the quantized transmission originating from the 14-AGNR band structure in (a). (c) \Inter-GNR transmissions $T_{14}$, $T_{23}$, $T_{24}$ and $T_{13}$ as a function of an \inter-ribbon voltage $V_{14}$ ($eV_{14}\equiv\mu_1-\mu_4$, with $\mu_1=\mu_2$ and $\mu_3=\mu_4$). 
}
\label{fig:trans90}
\end{figure*}

We now start analyzing the effect of the scattering due to the interaction between the ribbons both in the \intra- and \inter-GNR transport characteristics. 
In \Figref{fig:trans90}b we present the \intra-GNR transmissions $T_{12}$ (blue) and $T_{34}$ (dashed red line) obtained for the $\theta=90^\circ$ structure as a function of a \inter-GNR voltage $V_{14}$, \ie, the bias is applied so to create a potential difference between the two GNRs ($eV_{14}\equiv\mu_1-\mu_4$, with $\mu_1=\mu_2$ and $\mu_3=\mu_4$). As mentioned above, the zero-bias transmission of a pristine 14-AGNR (dashed green) serves as a reference.

The main effect observed in the \intra-GNR transmissions is a rigid shift of the ribbons' electronic levels by $\pm e V_{14}/2$.
At higher energies, far from $E_F$, dips on the \intra-GNR transmission are observed, which are related to the increase of the backscattering probability, or reflection function, $R_j(E)=N_T(E)-\sum_{i\neq j}T_{ji}(E)$ (for electrode 1 see orange dashed curves at \Figref{fig:trans90}b).

For the $\theta=90^\circ$ device the \intra-GNR transmission is considerably larger than the \inter-GNR, as can be seen in \Figref{fig:trans90}c (notice the logarithmic scale in this figure, see also the top panels in \Figref{fig:transRot}).
With the increasing of bias, the major effect in the \inter-GNR is a widening of the transmission gap around $E_F$ (\Figref{fig:trans90}c), which is proportional to the energy difference between the position of the CB of GNR$_{12}$ (whose levels were shifted up in energy by the applied bias) and that of the VB from GNR$_{34}$ (whose levels were shifted down in energy). 
When the applied bias achieves the same order of magnitude as the energy gap ($E_g=132$ meV), the VB from GNR$_{12}$ reaches the CB of GNR$_{34}$, which gives rise to an \inter-GNR transmission at $E_F$, as shown in \Figref{fig:trans90}c for $V_{14}=0.15$ V. For higher bias, \eg, $V_{14}=0.5$ V, the overlap between the GNR$_{12}$ VB and the GNR$_{34}$ CB increases and, as a result, an \inter-GNR transmission plateau is formed  around $E_F$ that widens with the applied voltage.

The \inter-GNR transmissions $T_{14}$ and $T_{23}$ (as well as $T_{13}$ and $T_{24}$) exhibit a very similar behavior, which is due to the high degree of symmetry of the system. This symmetry becomes even more evident for devices with $\theta \neq 90^\circ$ and, therefore, we show only the \inter-GNR transmissions $T_{13}$ and $T_{14}$ from here on. Note, however, that they are not exactly equivalent because the 14-AGNR (see inset to \Figref{fig:trans90}a) does not possess mirror symmetry along the axis defining its extended direction.

\subsection{Role of intersection angle}

\begin{figure*} 
\includegraphics[width=\textwidth]{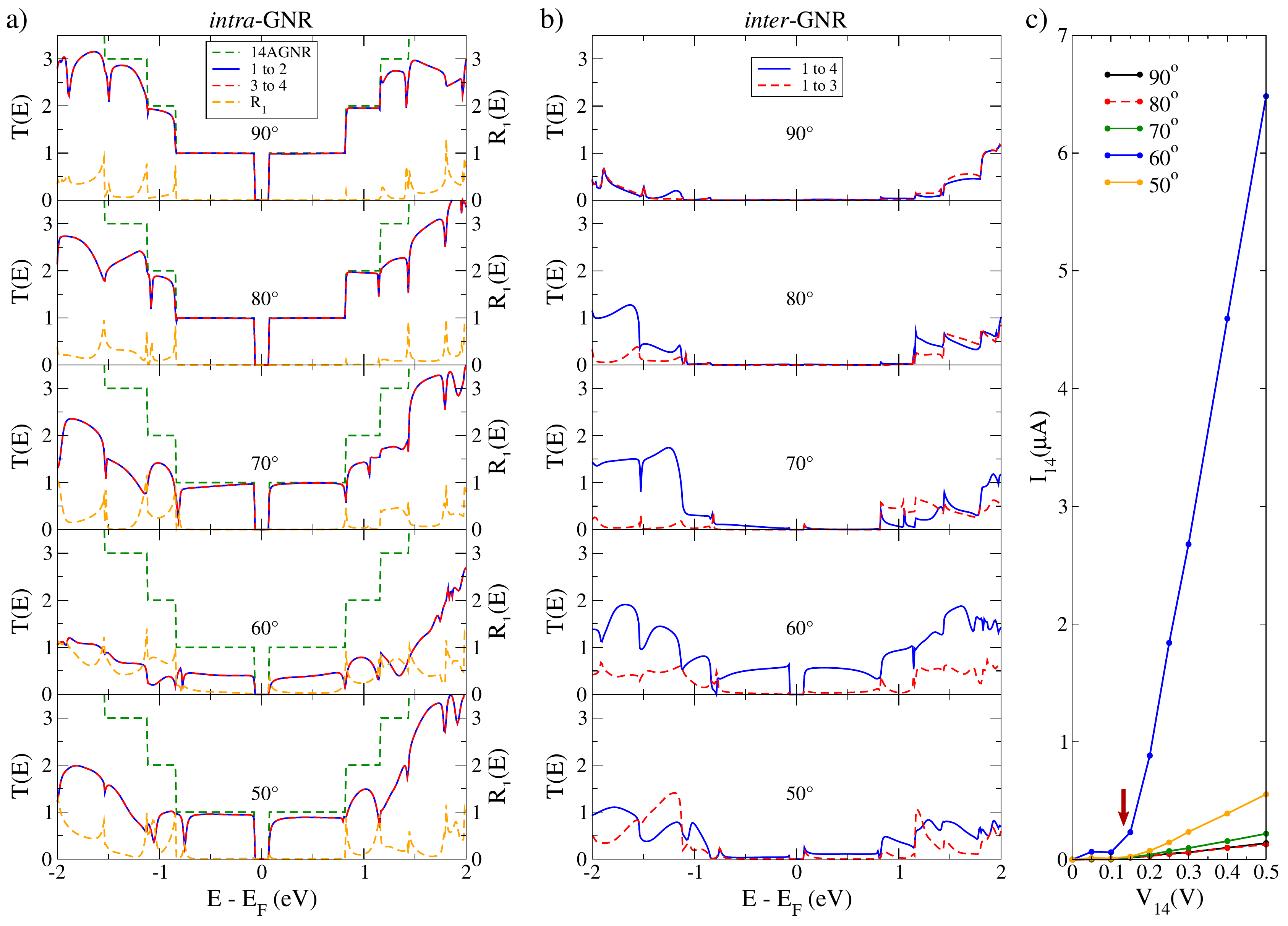}
\caption{Role of intersection angle $\theta$ for the transport characteristics. 
(a) \Intra-GNR transmissions $T_{12}$ (blue) and $T_{34}$ (dashed red) and reflection $R_1$ (dashed orange)  at zero bias ($\mu_1=\mu_2=\mu_3=\mu_4$).
Green dashed curves show the quantized transmission originating from the 14-AGNR band structure (\Figref{fig:trans90}a).
(b) \Inter-GNR transmissions $T_{14}$ (blue) and $T_{13}$ (dashed red) at zero bias ($\mu_1=\mu_2=\mu_3=\mu_4$).
(c) Current $I_{14}$ flowing from electrode 1 to 4 as a function of an \inter-GNR voltage $V_{14}$ ($eV_{14}\equiv\mu_1-\mu_4$, with $\mu_1=\mu_2$ and $\mu_3=\mu_4$).
A red arrow indicates the onset of \inter-GNR current at $V_{14} \sim E_g=132$ meV.}
\label{fig:transRot}
\end{figure*}

An interesting phenomenon is discovered when one varies the intersection angle $\theta$ between the GNRs.
At \Figref{fig:transRot}a we show the zero bias \intra-GNR transmissions $T_{12}$ and $T_{34}$ calculated for different angles $\theta=90^\circ$, $80^\circ$, $70^\circ$, $60^\circ$ and $50^\circ$.
Again the pristine 14-AGNR transmission (dashed green) is included for reference. Essentially, one observes an overall reduction of the \intra-GNR transmission with the decrease of $\theta$.
The lowest transmission values close to $E_F$ were obtained with the $\theta=60^\circ$ structure, exactly where one finds a closer matching between the honeycomb lattice of both ribbons in the crossing region. 
This decrease of the \intra-GNR transmission with $\theta$ also translates into the opposite behavior of the \inter-GNR transmission $T_{14}$, which tends to increase (\Figref{fig:transRot}b). 
The effect is particularly dramatic for the $\theta=60^\circ$ case, where one finds that $\sim 50\%$ of one \inter-GNR transmission channel is open in the energy window $E-E_F\in[-0.8,0.8]$~eV.
Surprisingly, the devices with the closer angles among those studied ($\theta=70^\circ$ and $\theta=50^\circ$) exhibit a T$_{14}$ that is at least one order of magnitude smaller in the mentioned energy range.
An additional interesting observation is that the device with $\theta=70^\circ$ exhibits a larger \inter-GNR transmission $T_{14}$ for the VB than for the CB, while for $\theta=50^\circ$ the situation is reversed.

One important property observed for all considered rotation angles is the low reflection probability around the Fermi energy (see for instance the electrode 1 reflection function $R_1(E)$ at \Figref{fig:transRot}a), indicating that in absence of external potential low energy electrons can propagate with negligible backscattering.

In \Figref{fig:transRot}c we present the calculated current $I_{14}$ flowing from electrode 1 to 4 as a function of an \inter-GNR applied bias $V_{14}$. The $60^\circ$ structure stands out when compared to all other cases, showing an \inter-GNR current $I_{14}$ higher by one order of magnitude, in accordance with what one would expect from the zero bias transmission analysis. The red arrow at \Figref{fig:transRot}c indicates the onset of the \inter-GNR current at $\sim E_g=132$ meV (the non-zero values of $I_{14}$ bellow the onset observed for 60$^\circ$ are attributed to the small broadening used in the calculations).

At $\theta=60^\circ$, the \inter-GNR transport proves to be more than just a secondary effect. Rather it is as significant as the direct \intra-GNR transport. Moreover, comparing the \inter-GNR transmissions $T_{13}$ and $T_{14}$ (\Figref{fig:transRot}b), one can predict that the scattering states from electrode 1 that are transmitted to the crossing GNR will propagate most likely towards the electrode 4 rather than 3 for all  $\theta < 90^\circ$, a remark that is most evident for $\theta=60^\circ$.

\begin{figure*} 
\includegraphics[width=\textwidth]{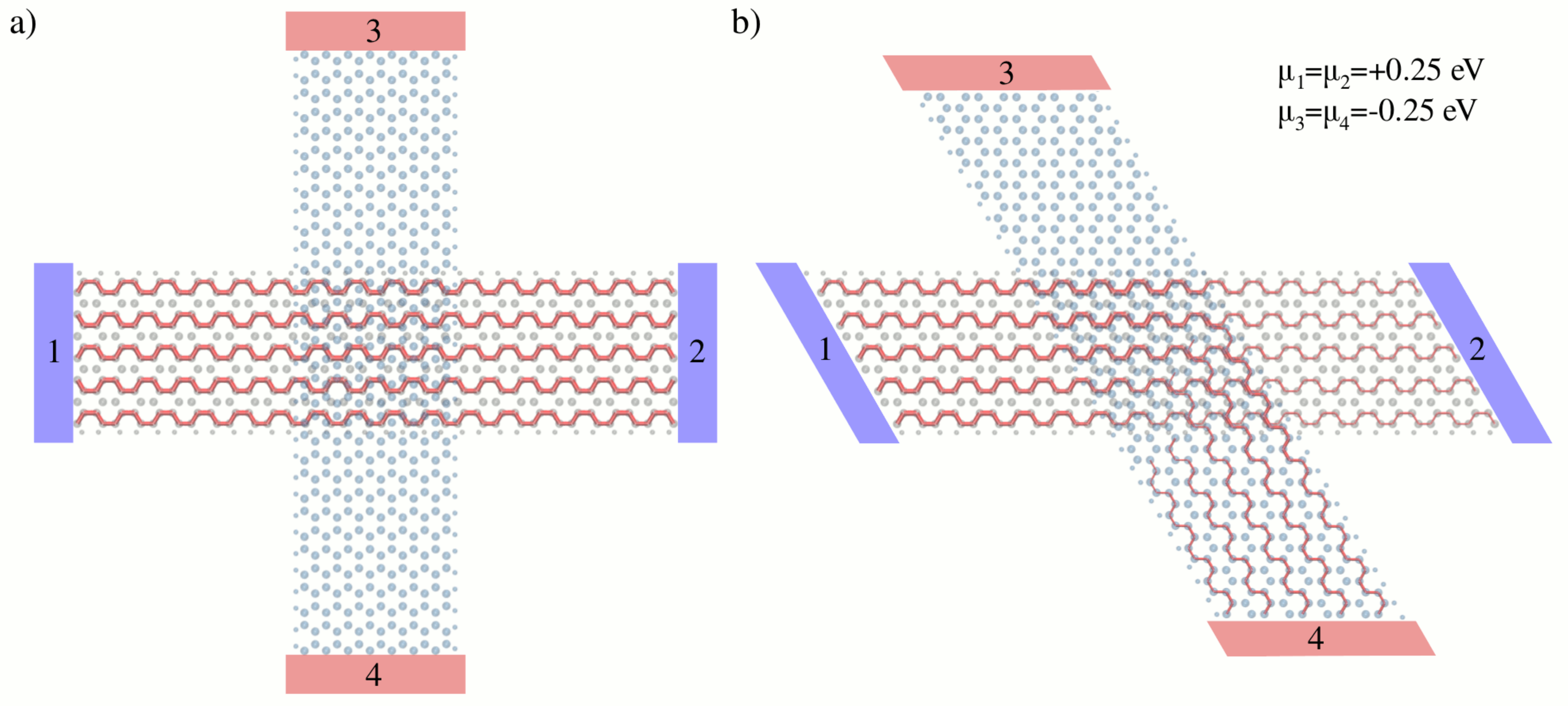}
\caption{Bond currents generated from electrode 1 scattering states integrated over the energy interval $\left|E-E_F\right|<0.5$ eV under an \inter-GNR voltage of $V_{14}=0.5$ V ($eV_{14}\equiv\mu_1-\mu_4$, with $\mu_1=\mu_2$ and $\mu_3=\mu_4$) for (a) $\theta=90^\circ$ and (b) $\theta=60^\circ$ devices. The contacts to the electrodes are shown with colored boxes representing the applied bias.}
\label{fig:bondcurr}
\end{figure*}

This prognosis is confirmed with the bond currents from electrode 1 calculated with an \inter-GNR voltage of $V_{14}=0.5$ V ($eV_{14}\equiv\mu_1-\mu_4$, with $\mu_1=\mu_2$ and $\mu_3=\mu_4$) and integrated over the energy window $\left|E-E_F\right|<0.5$ eV.
On the one hand, with a $90^\circ$ setup (\Figref{fig:bondcurr}a), 
all scattering states from electrode 1 almost fully propagate towards terminal 2 and essentially no current flows to the crossing ribbon.\footnote{For a better visualization of the bond currents a cutoff was applied so that all bond currents below a given threshold are not printed.}  On the other hand, for $\theta = 60^\circ$ (\Figref{fig:bondcurr}b) only about half of the states propagates towards terminal 2, while the other half is transmitted through the crossing to electrode 4, and no current flows from 1 to 3.

\subsection{Operation of one GNR as a gate electrode}

In this section we study the transport characteristics in the crossing as a function of \inter- and \intra-GNR voltages. The \intra-GNR voltage $V_{12}$ was applied only among the electrodes 1 and 2, \ie, $eV_{12}\equiv\mu_1-\mu_2$, while the electrodes 3 and 4 were maintained at the same chemical potential, $\mu_3=\mu_4$. The \inter-GNR voltage $V_{14}$  was defined by the difference between the chemical potentials from electrode 1 and 4, \ie, $eV_{14}\equiv\mu_1-\mu_4$. Therefore, within this setup one can investigate how the GNR$_{34}$ can act as a gate to the current flowing through GNR$_{12}$ in crossing systems presenting low \inter-GNR transmission, such as $\theta=90^\circ$. Moreover, this allows one to tune the current splitting on devices with higher \inter-GNR transmission, which is the case for $\theta = 60^\circ$.

\begin{figure} 
\includegraphics[width=.48\textwidth]{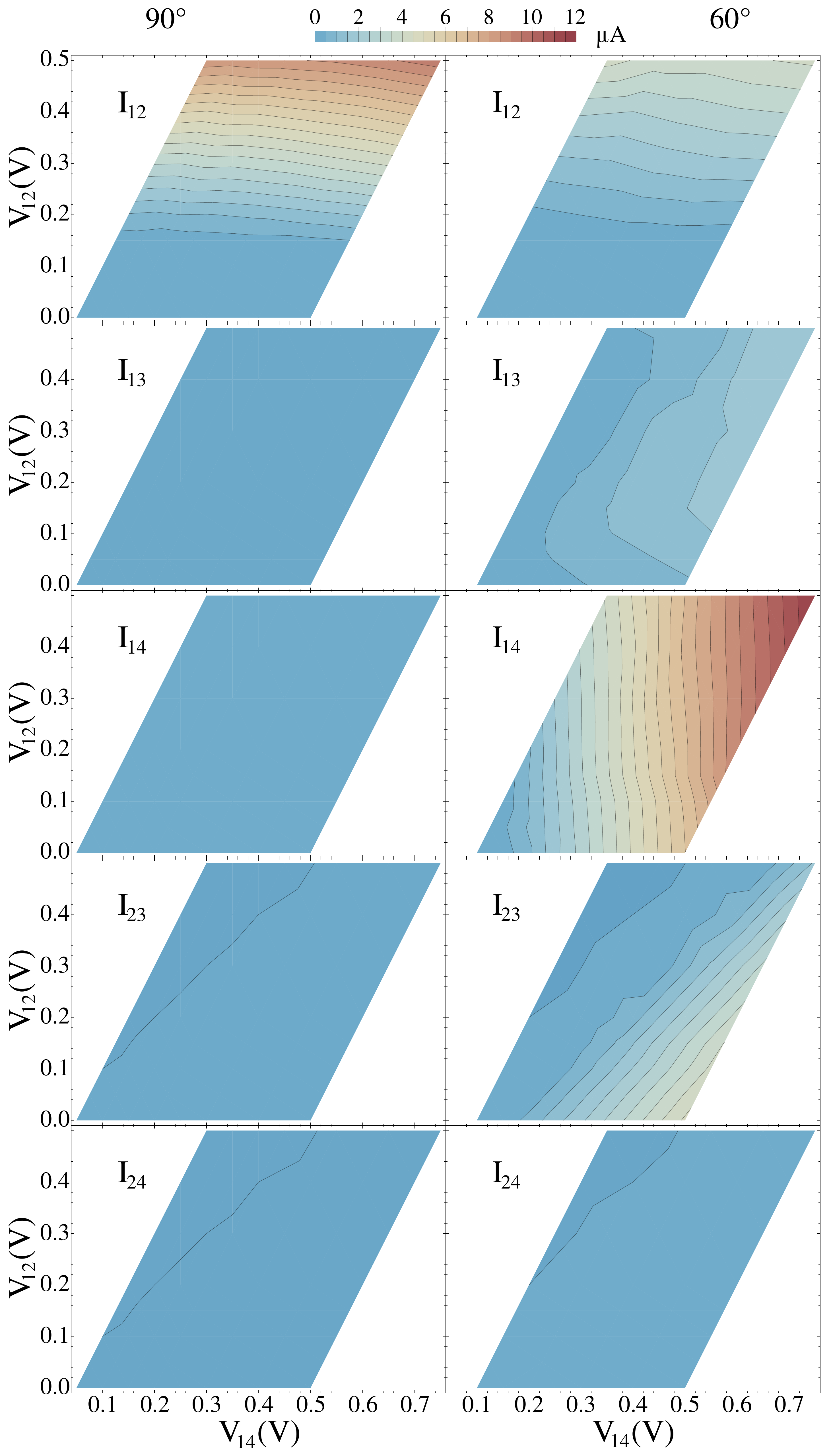}
\caption{Contour plots of \intra-GNR currents ($I_{12}$) and \inter-GNR currents ($I_{13}$, $I_{14}$, $I_{23}$ and $I_{24}$) as a function of the \intra-GNR ($eV_{12}\equiv \mu_1-\mu_2$) and of the \inter-GNR ($eV_{14} \equiv \mu_1-\mu_4$) voltages, with the electrodes 3 and 4 maintained at the same chemical potential ($\mu_3=\mu_4$). The results for the 90$^\circ$ device are displayed on the left and the corresponding for 60$^\circ$ on the right. All plots share the same color scale shown on top.}
\label{fig:intra}
\end{figure}

In \Figref{fig:intra} we present the different components for the current with the variation of $V_{12}$ and $V_{14}$, for both $\theta=90^\circ$ and $\theta=60^\circ$ devices.
The \intra-GNR current $I_{12}$ presents an onset at $\sim E_g=132$ meV and is more sensitive for $90^\circ$ (top left panel in \Figref{fig:intra}) where it clearly increases fast with $V_{12}$ but slower with $V_{14}$, indicating that the GNR$_{34}$ produces only a weak gating effect on the current flowing through GNR$_{12}$.
We note that this weak gating effect is in contrast with the calculations reported in Ref.~\onlinecite{Habib2012}, showing a current variation of several orders of magnitude with the \inter-GNR bias.
A possible reason for this discrepancy could be related to the nonequilibrium charge redistribution, an effect which we include in our present study.
The \inter-GNR current components $I_{13}$, $I_{14}$, $I_{23}$ and $I_{24}$ for 90$^\circ$ (left panels in \Figref{fig:intra}) are all negligible compared to the \intra-GNR $I_{12}$, and essentially no change is observed within the applied bias range.

For the $\theta=60^\circ$ device (right panels in \Figref{fig:intra}) the \inter-GNR currents $I_{14}$ and $I_{23}$ present the same order of magnitude for $V_{12}=0$. When a finite \intra-GNR voltage is applied the main effect observed is that the current flowing to GNR$_{34}$ arises more from electrode 1 and less from 2, meaning that the electron splitting can be tuned combining $V_{12}$ and $V_{14}$.

\subsection{Analysis of the scattering potential at the crossing}

\begin{figure*} 
\includegraphics[width=\textwidth]{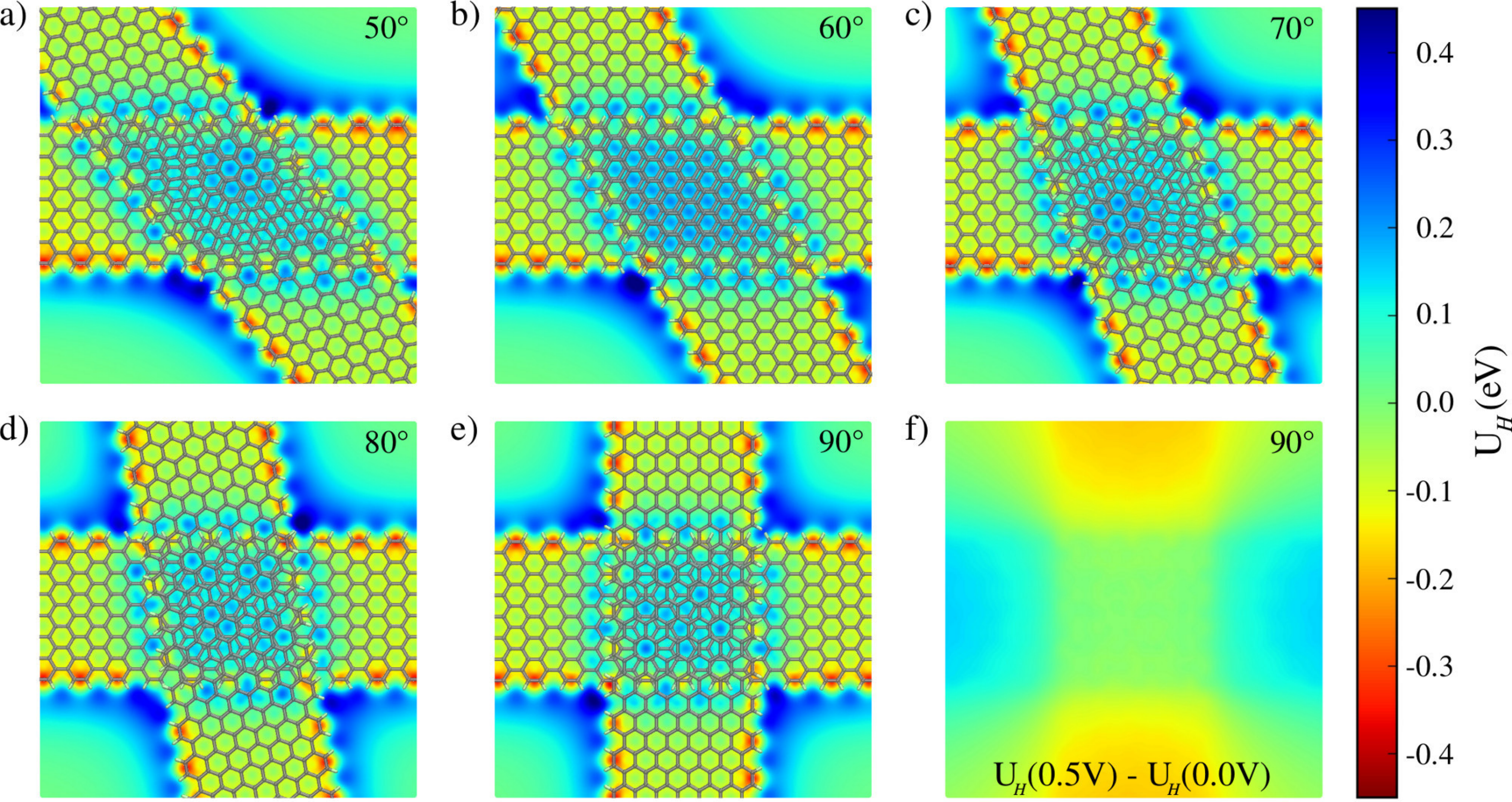}
\caption{Electrostatic potential $U_H$, \ie, the Hartree potential obtained from the self-consistent density plus local pseudopotential, visualized at the intermediary plane between the GNRs with no applied voltage ($\mu_1=\mu_2=\mu_3=\mu_4$) for devices with different intersection angles $\theta$: (a) 50$^\circ$, (b) 60$^\circ$, (c) 70$^\circ$, (d) 80$^\circ$, and (e) 90$^\circ$. The representations of the molecular structures were superimposed to the plots to guide the visualization. (f) The electrostatic potential \emph{difference} for an \inter-GNR voltage of $V_{14} = 0.5$ V ($eV_{14}\equiv\mu_1-\mu_4$, here with $\mu_1=\mu_2$ and $\mu_3=\mu_4$) with respect to zero voltage for the $\theta=90^\circ$ device. All plots share the same color scale.}
\label{fig:hartree}
\end{figure*}

In order to characterize the change of the scattering properties in the crossing region, we present here results for the distribution of the electrostatic potential  in the central region. This is defined as the Hartree potential plus the local pseudopotential describing the electron-ion interaction within the \textsc{Siesta}/\textsc{TranSiesta} packages~\cite{Soler2002,Brandbyge2002,Papior2016}. \Figref{fig:hartree}a-e show the electrostatic potential at the middle plane between the two ribbons resulting from our calculations at different angles $\theta$ and without applied voltage ($\mu_1=\mu_2=\mu_3=\mu_4$). 
For all angles the potential is higher (more repulsive for electrons) in the crossing region, and with the highest values in regions where the lattices of the two ribbons match. 
This stems from the electron charge accumulation in the \inter-ribbon region.
Accordingly, for the $\theta=60^\circ$ case, where the lattices happen to match within the entire intersection region, the potential reveals ``bumps'' over the entire crossing, which might be interpreted as a source of larger scattering (and, thus, a harder barrier) for propagating electrons.
Thus, one might be tempted to assign to this larger corrugation of the effective electron potential the simultaneous decrease of the \intra-GNR and increase of the \inter-GNR scattering at $\theta=60^\circ$. 

\Figref{fig:hartree}f explores whether this effect can be strongly modified at finite bias. In this figure we show the difference of the electrostatic potential for an \inter-GNR voltage $V_{14}=0.5$ V ($eV_{14}\equiv\mu_1-\mu_4$, with $\mu_1=\mu_2$ and $\mu_3=\mu_4$) and a zero bias ($\mu_1=\mu_2=\mu_3=\mu_4$) calculations for the $90^\circ$ device. This plot only reveals smooth changes in the self-consistent electrostatic potential due to the applied bias. In particular, we do not find noticiable changes in the crossing region. This indicates that the electron scattering at the crossing will not be drastically modified by the \inter-GNR bias, in agreement with the general trends observed for the transmission functions presented so far. 

\subsection{Role of inter-ribbon distance}

The above analysis of the electrostatic potential in the crossing suggests that the overlap of carbon $\pi$-orbitals may produce a strong effect on the potential distribution and, thus, on the scattering at the crosssing, increasing the \inter-GNR transmission. In order to test this hypothesis and to have a better understanding of the observed transport properties, we now consider the role of the inter-ribbon distance on the ratio between the \intra- and \inter-GNR transport. 

\begin{figure} 
\includegraphics[width=.48\textwidth]{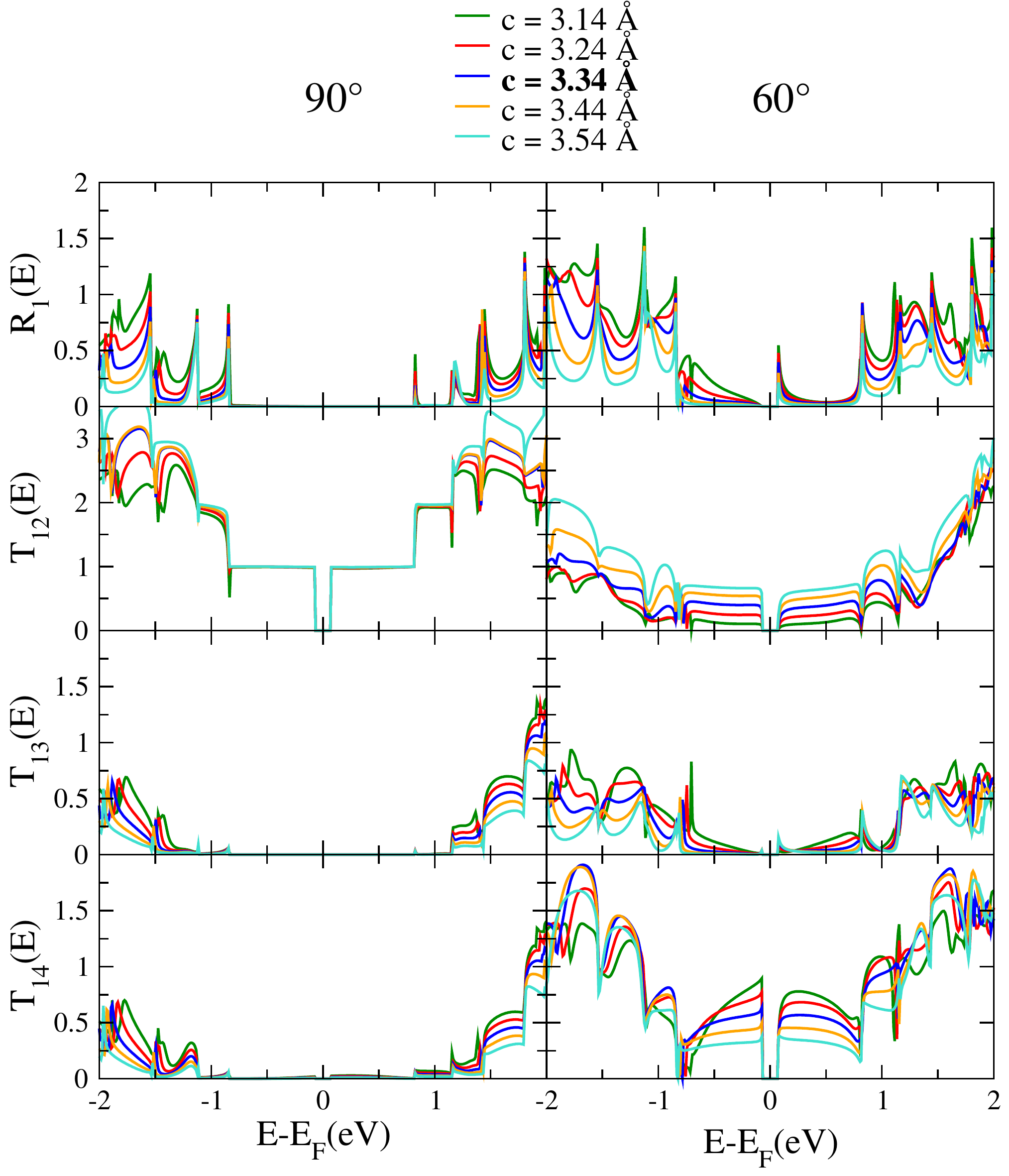}
\caption{Reflection $R_1$ and transmissions $T_{12}$, $T_{13}$, and $T_{14}$ for $\theta=90^\circ$ (left) and $\theta=60^\circ$ (right) for different \inter-GNR distances $c$. 
The structures obtained by rotating the duplicated GNR with respect to its center of mass, \ie, those discussed up to now, are highlighted with thick lines.
No voltage is applied ($\mu_1=\mu_2=\mu_3=\mu_4$).}
\label{fig:height}
\end{figure}

In \Figref{fig:height} it is presented the electrode 1 zero bias reflection and transmissions as a function of the inter-ribbon distance around our reference value $c=3.34$ \AA. We analyze two extreme cases, namely $\theta=90^\circ$ (the case with higher \intra-GNR and lower \inter-GNR transmissions) and $\theta=60^\circ$ (with lower \intra-GNR and higher \inter-GNR transmissions).  
Inside the varying interval of $\pm 0.2$ \AA, almost no change is observed in the transmission for the 90$^\circ$ device close to $E_F$. 
For higher energies, $\left| E - E_F \right| > 1$ eV, the decreasing distance between the GNRs infers a stronger scattering effect, which is expressed in terms of the reflection probability $R_1$.
The dependence with the distance is significantly different for the 60$^\circ$ case (\Figref{fig:height}, on the right).
As the distance between GNRs decreases we observe a clear increase of the \inter-GNR transmission.
This takes place at the expense of the \intra-GNR transport, which gets drastically reduced.
We note that this result indicates that the transmission in a $\theta=60^\circ$ device could be tuned by applying an external force to the junction.
The feasibility of this kind of electromechanical switching has been also suggested for crossed carbon nanotubes.~\cite{Yoon2001} 

\section{Discussion}

\subsection{Lattice matching and registry index}
\begin{figure} 
\includegraphics[width=.48\textwidth]{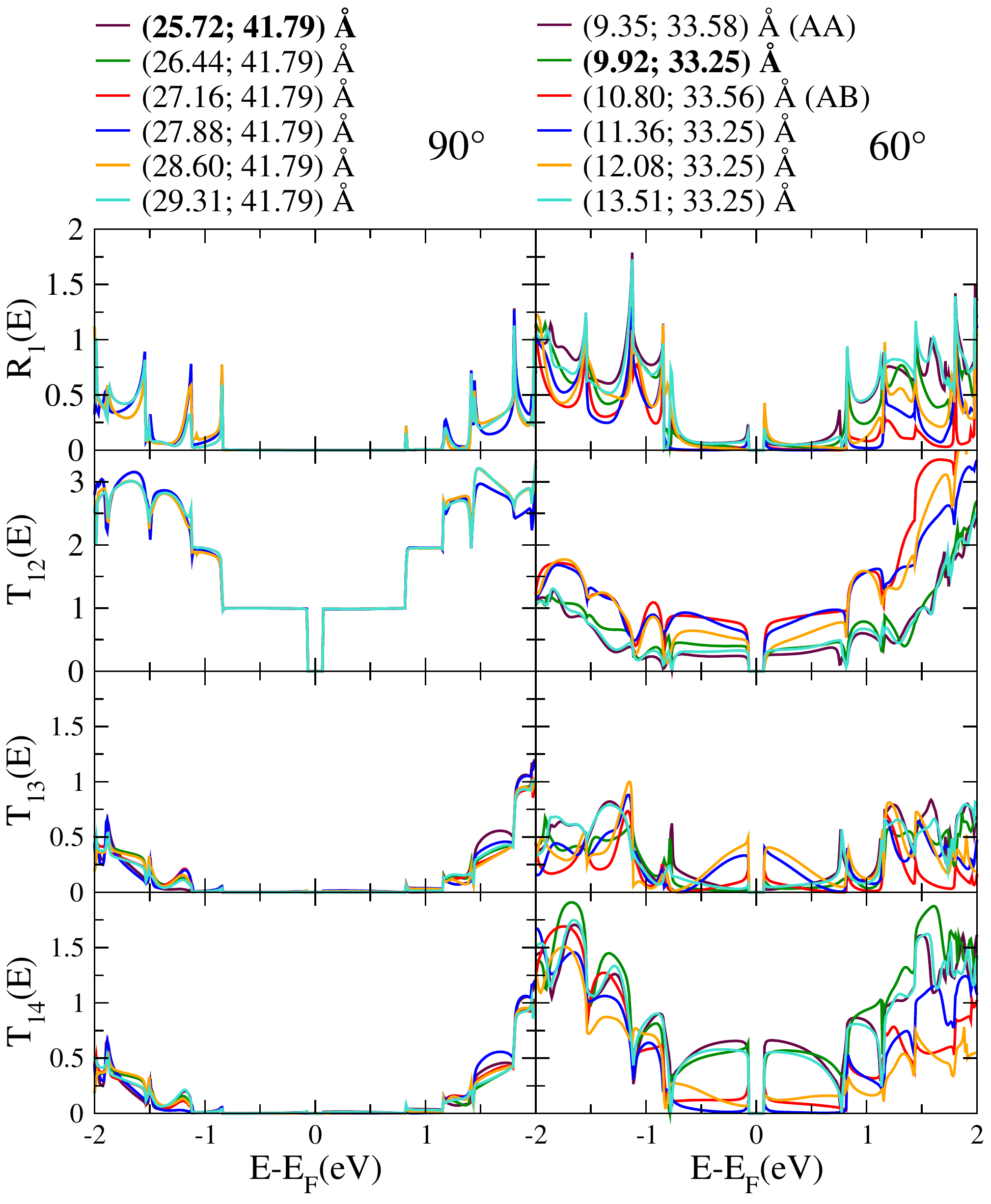}
\caption{Reflection $R_1$ and transmissions $T_{12}$, $T_{13}$, and $T_{14}$ for $\theta=90^\circ$ (left) and $\theta=60^\circ$ (right) as a function of horizontal translation of one ribbon with respect to the other. The \inter-GNR distance is fixed at $c=3.34$ {\AA} and no voltage is applied ($\mu_1=\mu_2=\mu_3=\mu_4$).
The structures obtained by rotating the duplicated GNR with respect to its center of mass, \ie, those discussed up to now, are highlighted with thick lines.}
\label{fig:translate}
\end{figure}

To investigate the role of lattices matching in the crossing region, another set of calculations was performed by translating one GNR with respect to the other while keeping the \inter-GNR distance at $c = 3.34$ \AA.
In \Figref{fig:translate} the electrode 1 zero bias reflection and transmissions for structures with $\theta = 90^\circ$ and $\theta = 60^\circ$ are presented for the six different stacking configurations each.
Very little difference is observed in the transmissions among the translated structures with 90$^\circ$ (left panels in \Figref{fig:translate}).
This is consistent with the idea of the overlap between the $\pi$ orbitals being the key parameter, since the average overlap does not vary much when translating the GNRs with $\theta = 90^\circ$.
In other words, an average of the different stackings between carbon atoms in the two ribbons is always sampled when the ribbons cross at $\theta = 90^\circ$ and, therefore, a small shift of the ribbons' positions does not qualitatively change the situation.

In contrast, for the 60$^\circ$ case the results show a strong change already close to $E_F$ (right panels in \Figref{fig:translate}).
Two particular stackings can be highlighted: AA (where the carbons from the different ribbons lay on top of each other in the crossing region) and AB (when half of the carbons lay on top of other carbons, while half of them resides on the center of the other GNR hexagons and, therefore, does not overlap).
These two stacking correspond exactly to the maximum (AA) and minimum (AB) possible overlap between the carbon atoms in the crossing.
Accordingly, the AA case presents the highest/lowest \inter-/\intra-GNR transmission, while for AB stacking one find the lowest/highest \inter-/\intra-GNR transmission.

So far, all results support the hypothesis that the overlap between $\pi$-orbitals in the crossing determines the scattering properties in our system. However, if this simple picture would be enough to explain all the observed phenomena, one could in principle quantify the amount of scattering using some measure that characterizes the overlap for each structure. The registry index (RI) can be used to provide such a measure (see \cite{Hod2013} for a detailed review). The idea is to consider a circle around each carbon atom belonging to the crossing region and compute the overlapping area $S_\mathrm{CC}$ between the circles from different GNRs. For graphene like materials it has been shown that the ideal circle radius to be considered corresponds to half of a C-C covalent bond length in graphene (\ie, 0.71 \AA). The RI is then defined as $\mathrm{RI} = \left(S_\mathrm{CC}-S_\mathrm{AB}\right)/\left(S_\mathrm{AA}-S_\mathrm{AB}\right)$, where $S_\mathrm{AA}$ and $S_\mathrm{AB}$ are respectively the maximum and minimum possible overlaps between the GNR orbitals.

Table~\ref{tab:ri} shows the RI calculated for structures with different intersection angles together with the special cases of $\theta = 60^\circ$ with AA and AB stacking. Considering all the atoms in the crossing region (RI$_\mathrm{tot}$), the highest value is obtained for $\theta = 60^\circ$ with AA stacking (RI$_\mathrm{tot}=1.0$) and the minimum one for $\theta = 60^\circ$ with AB stacking (RI$_\mathrm{tot}=0.0$), as one would expect from the definition above. This values qualitatively describes the changes of the \inter-GNR currents among the 60$^\circ$ cases [$I_{14}^\mathrm{AA}(0.5V)=7.96\mu$A versus $I_{14}^\mathrm{AB}(0.5V)=1.10\mu$A]. Moreover, the calculated RI$_\mathrm{tot}$ exhibits a qualitative agreement with the transport properties for $\theta=90^\circ$, $80^\circ$, $70^\circ$ and $50^\circ$ devices. However, a discrepancy occurs when all cases are contemplated, since all devices with $\theta = 60^\circ$, including the AB stacking with RI$_\mathrm{tot}=0.0$, present higher \inter-GNR current than all the other devices with different $\theta$, for which RI$_\mathrm{tot} > 0.0$.

\begin{table}[h]
  \begin{center}
    \begin{tabular}{l|c|c|c|c|c|c|c}
      \hline
      \hline
      device & $\:\;\;90^\circ\;\;$ & $\:\;\;80^\circ\;\;$ & $\:\;\;70^\circ\;\;$ & $\:\;\;60^\circ\;\;$ & $\:\;\;50^\circ\;\;$ & 60$^\circ$AA & 60$^\circ$AB \\
      \hline
      RI$_\mathrm{tot}$  & 0.08 & 0.11 & 0.17 & 0.45 & 0.44 & 1.00 & 0.00 \\
      RI$_\mathrm{edge}$ & 0.00 & 0.24 & 0.35 & 0.12 & 0.55 & 1.00 & 0.15 \\
      \hline
      \hline
    \end{tabular}
    \caption{Registry index computed for devices with different intersection angles and stacking. RI$_\mathrm{tot}$ column shows the values obtained when all the carbons have been considered, whereas for RI$_\mathrm{edge}$ only carbons belonging to the edges of the GNRs were taken into account.}
    \label{tab:ri}
  \end{center}
\end{table}

One could consider that, for example, only the GNR edges in the crossing are relevant for describing the scattering properties. Hence the registry index can be calculated only considering the overlaps from carbons belonging to the GNR edges (RI$_\mathrm{edge}$). This will change the RI to quantitative different values, see Table~\ref{tab:ri}. However, none of the registry index values does qualitatively describe the angle dependency on the current when all cases are taken into account.

\subsection{Simple model for \inter-ribbon tunneling}
The analysis in the previous section shows that the overlap of $\pi$ orbitals in the crossing region cannot alone account for the observed \inter-GNR transmission and, in particular, explain what makes a device with $\theta=60^\circ$ a very interesting and effective candidate as a beam splitter.

\begin{figure*} 
\includegraphics[width=\textwidth]{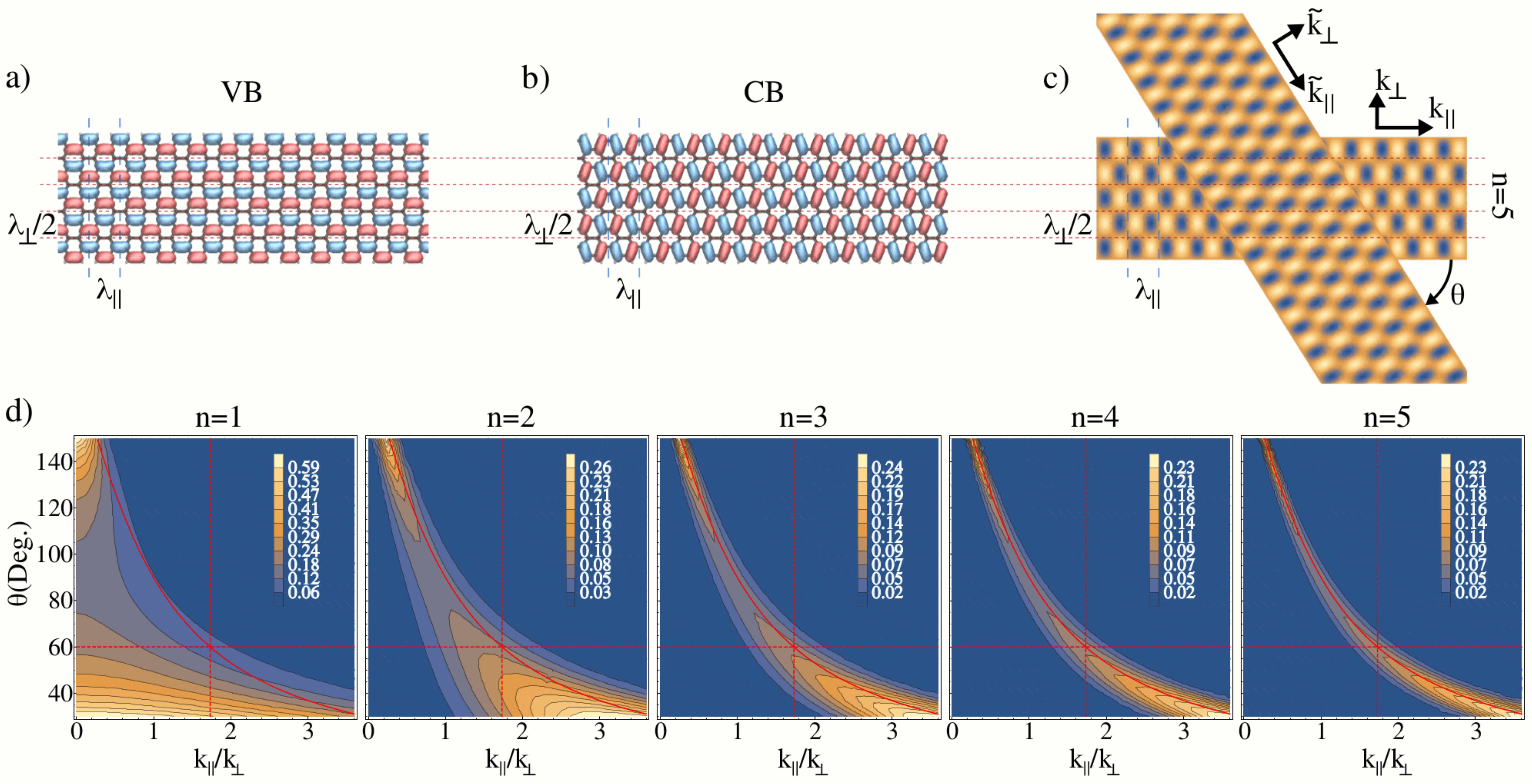}
\caption{(a) CB and (b) VB wavefunctions of a 14-AGNR calculated at the $\Gamma$ point with DFT, both revealing a wave length ratio of $\lambda_\perp/\lambda_\parallel=\sqrt{3}$ as expected for electrons in graphene sufficiently near the Dirac point. (c) Density plot of the real part of the qualitatively similar wavefunction of \Eqref{eq:wf} with four nodal planes ($n=5$) across the GNR. (d)~Contour plots of the probability amplitude $\big| \langle \Psi_{\tilde{\mathbf{k}}_\parallel,\tilde{\mathbf{k}}_\perp} | \Psi_{\mathbf{k}_\parallel,\mathbf{k}_\perp} \rangle\big|^2$ evaluated numerically for the first five fundamental modes $n$ as a function of the ratio $k_\parallel/k_\perp$. The region with significant overlap (orange area) is observed to narrow in as $n$ increases. The full red curve represents the wavenumber matching condition from \Eqref{eq:maxoverlap} (\ie, maximum overlap in the large $n$ limit), which at $k_\parallel/k_\perp=\sqrt{3}$ corresponds to exactly $\theta=60^\circ$ (red dashed lines).}
\label{fig:wfoverlap}
\end{figure*}

The key to understand the physical origin of this important effect is to consider the tunneling probability between the relevant electronic states in each of the two crossing GNRs. To do this properly it is necessary to take into account not only the overlaps between atomic orbitals in neighboring structures, but also the relative phases and amplitudes with which these orbitals participate to those scattering states. The VB and CB states of the 14-AGNR calculated at the $\Gamma$ point with DFT are shown in \Figref{fig:wfoverlap}a-b. 
These two states are representative for the available electron bands in the energy range of $\pm 0.8$ eV around $E_F$.
While the VB is characterized by an odd symmetry with respect to the GNR axis and the CB by an even symmetry, both states share a common structure with four nodal planes along the ribbon.
Reminiscent of the electron states at the Dirac point in graphene,\cite{CastroNeto2009} these states reflect the characteristic ratio
\begin{align}
\frac{\lambda_\perp}{\lambda_\parallel}=\frac{k_\parallel}{k_\perp}=\sqrt{3}
\quad\mathrm{(for\; AGNRs)},
\label{eq:perp-par-ration}
\end{align}
between the electron momentum along the ribbon axis ($k_\parallel$) and perpendicular to it ($k_\perp$). 
The AGNR states can be interpreted as quantized states of a graphene layer  and, thus, they can be qualitatively represented by a propagating wave along the ribbon axis with a Bloch wavevector $\mathbf{k}_\parallel$ together with a particle-in-a-box state corresponding to a wavevector $\pm\mathbf{k}_\perp$ in the confined (perpendicular) direction.~\cite{Wakabayashi2010} Ignoring the details of the wavefunction inside the graphene unit cell, and just focusing on the envelope wavefunction, we just approximate this situation using plane-waves, in which case we are left with the traditional bands of a free-electron wire, \ie, we have
\begin{equation}
\langle \mathbf{r}|\Psi_{\mathbf{k}_\parallel,\mathbf{k}_\perp} \rangle
= \left\{  
\begin{array}{l}
	e^{-i \mathbf{k}_\parallel \cdot \mathbf{r}} \big(e^{-i \mathbf{k}_\perp \cdot \mathbf{r}}-e^{i \mathbf{k}_\perp \cdot \mathbf{r}}\big),\; \mathbf{r} \in \mathrm{GNR} \\
0,\quad \mathrm{elsewhere}
\end{array}\right.
\label{eq:wf}
\end{equation}
where $k_\perp = n \pi /W$ depends on the GNR width $W$ and the quantum number $n$ (positive integer). 
The second AGNR is described similarly, but with rotated wavefunctions $|\Psi_{\tilde{\mathbf{k}}_\parallel,\tilde{\mathbf{k}}_\perp} \rangle$ where the wavevectors in the two ribbons are related \textit{via} the rotation matrix $\mathbf{R}(\theta)$ defined for a clockwise rotation angle $\theta$:
\begin{align}
\tilde{\mathbf{k}}_{\parallel} &= \mathbf{R}(\theta) \mathbf{k}_{\parallel},\\
\tilde{\mathbf{k}}_{\perp} &= \mathbf{R}(\theta) \mathbf{k}_{\perp}.
\label{eq:rotationmatrix}
\end{align}
In an elastic scattering process the energy is conserved, \ie, the wavevectors generally fulfill the condition
\begin{align}
\tilde{k}_{\parallel} + \tilde{k}_{\perp}  &= {k}_{\parallel} + k_{\perp}.
\label{eq:energy-cons}
\end{align}
As can be seen in \Figref{fig:wfoverlap}c, the electronic structure obtained for the AGNRs using this simple particle-in-a-box quantization condition is qualitatively correct.
In the spirit of perturbation theory, the \inter-GNR tunneling probability \cite{Bardeen1961,Habib2012,VanDePut2016} is assumed to be proportional to the modulus square of the overlap between the two wavefunctions,
\begin{equation}
T_\mathit{inter} \propto \big|\langle\Psi_{\tilde{\mathbf{k}}_\parallel,\tilde{\mathbf{k}}_\perp} | \Psi_{\mathbf{k}_\parallel,\mathbf{k}_\perp} \rangle\big|^2.
\end{equation}
These overlaps can readily be evaluated numerically as shown in \Figref{fig:wfoverlap}d as a function of $n=\widetilde{n}$, $\theta$, and the ratio $k_\parallel/k_\perp$.
For the fundamental mode $n=1$ the two wavefunctions have a significant overlap in a large part of the parameter space.
In the limit of $\theta\rightarrow 0$ ($\theta\rightarrow \pi$), where the two GNRs are aligned in parallel (antiparallel), the overlap goes to infinity because of the diverging integration area.
As $n$ increases, the region with a significant overlap shrinks towards one universal curve.
This situation corresponds to the wavevector matching condition~\footnote{Note that another matching condition is given by the relation $\mathbf{k}_\parallel-\mathbf{k}_\perp =-\tilde{\mathbf{k}}_\parallel-\tilde{\mathbf{k}}_\perp$, from where one obtains a maximum for $\theta^*=120^\circ$, which taking into account the minus sign from $\mathbf{k}_\parallel$ corresponds to the same preferential scattering at 60$^\circ$.}
\begin{align}
\mathbf{k}_\parallel-\mathbf{k}_\perp &=\tilde{\mathbf{k}}_\parallel+\tilde{\mathbf{k}}_\perp,
\label{eq:matching} 
\end{align}
which in turn yields the relationship
\begin{align}
\cos\theta^* &= \frac{k_\parallel \tilde{k}_\parallel-k_\perp\tilde{k}_\perp}{k_\parallel^2+k_\perp^2}.
\label{eq:maxoverlap}
\end{align}
According to \Eqref{eq:perp-par-ration} this simply corresponds to $\theta^*=60^\circ$, \ie, the exact condition for a maximal \inter-GNR tunneling as found in our simulations for AGNR.
The meaning of the condition \Eqref{eq:matching} can be rephrased in very simple terms: for two ribbons interacting weakly and with a relatively large contact area (in units of the Fermi wavelength square) the tunneling probability will be maximized when the total wavevector of the electron is preserved in the elastic scattering process.

Notice that this simplified model cannot account for the dependence of the current on the stacking of the GNRs. In order to do so, in addition to the phases carried by the envelope wavefunctions it is necessary to account for the structure of the wavefunctions inside the graphene unit cell (and take into account for the overlaps between $\pi$ orbitals and their relative phases within the unit cell). 
This explains the partial success of the RI, that allows  rationalizing the changes of the \inter-GNR transport for a fixed angle $\theta$. However, the main effect of the rotation angle is accounted for  by our simplified model based on a description of the electronic states as plane waves.

Although \Eqref{eq:wf} can be a good approximation for the CB and VB of AGNRs, the situation is more complicated for nanoribbons of different orientations and, in particular, for ZGNRs.~\cite{Wakabayashi2009,CastroNeto2009,Wakabayashi2010,Lima2016}
While qualitatively correct at intermediate energies, the quantized graphene bands fail to describe important features of the low energy spectrum of ZGNRs, as the appearance of the edge states at the Fermi level.
Therefore, considering states sufficiently far from the Fermi energy, one can apply the simple model to ZGNRs, but in this case the relation between parallel and perpendicular momentum must be reversed,
\begin{align}
\frac{\lambda_\perp}{\lambda_\parallel}=\frac{k_\parallel}{k_\perp}=\frac{1}{\sqrt{3}}
\quad\mathrm{(for\; ZGNRs}).
\label{eq:perp-par-ration2}
\end{align}
Thus, for ZGNRs \Eqref{eq:maxoverlap} gives rise to a maximum tunneling probability for $\theta^*=120^\circ$, in agreement with the results reported in Ref.~\onlinecite{Lima2016} using a $\pi$-orbital tight-biding model of the system. 

Our results present clear connections with previous work investigating the modifications of bilayer graphene band structure as a function of the rotation angle of the two layers.~\cite{LopesdosSantos2007}
The above argumentation explains why the alignment of the honeycomb lattices of the AGNR ribbons at $60^\circ$ intersection radically increases the inter-ribbon interaction regardless of stacking, why this also happens for ZGNRs, although there the electron scattering takes place preferentially at 120$^\circ$.
It also explains the higher \inter-GNR conductance at 30$^\circ$ and 90$^\circ$ reported in Ref.~\onlinecite{Botello2011} for crossed AGNR/ZGNR devices.

\section{Conclusions}

In this paper we studied the electronic and transport properties of a 4-terminal junction defined by two crossed 14-AGNRs from \emph{first-principles} with the \textsc{Siesta}/\textsc{TranSiesta} codes~\cite{Soler2002,Brandbyge2002,Papior2016}.
Our research comprises a detailed investigation of the system behavior under the variation of structural parameters, such as intersection angle, \inter-GNR distance and stacking order, as well as its response under nonequilibrium conditions by considering different setups with a finite voltage applied between the electrodes.

Varying the intersection angle between the crossed AGNRs we found two extreme cases, namely $\theta =90^\circ$ and $\theta=60^\circ$ with low and high \inter-GNR transmission, respectively.
Remarkably, for the 60$^\circ$ case we found that the \inter-GNR transmission channel is close to $50\%$ and the reflection negligible over a relatively large energy window of $\pm0.8$ eV around the Fermi energy without an applied voltage.
Moreover, for all considered cases with $\theta<90^\circ$ the majority of \inter-transmitted electrons propagate only in one direction in the other ribbon.
Those findings indicate that semiconducting crossed AGNR structures are interesting candidates to be incorporated in quantum electronics devices.
In particular, we showed that a system with $\theta=60^\circ$ can operate as an electronic beam splitter where the ratio of \intra/\inter\ transmission can be tuned by changing the \inter-GNR distance, \ie, it can be mechanically controlled by applying an external force to the junction.

We also explored how the crossed structures behave with biased electrodes. Applying an \inter-GNR bias voltage, the 60$^\circ$ configuration is again distinguished with an \inter-GNR current one order of magnitude higher than all the other considered intersection angles.
When one AGNR is subjected to an \intra-GNR bias voltage, changing the \inter-GNR voltage produces a weak gating effect on the 90$^\circ$ devices, but reveals the possibility of tuning the current splitting on the 60$^\circ$ case.

Analyzing the electrostatic potential we found that the lattice matching on the crossing region plays an important role on the scattering properties.
Indeed, a significant change on the transmission probabilities is observed by varying the stacking order on the 60$^\circ$ device.
Those results suggest that the overlap of carbon $\pi$ orbitals is another essential parameter to the scattering process.
The structures' registry indices indicate that the amount of $\pi$-orbital overlap in the crossing can qualitatively describe the changes in the \inter-GNR currents among 60$^\circ$ cases as well as explain the trend in the transport properties for structures with $\theta \in \left\{ 50^\circ, 70^\circ, 80^\circ, 90^\circ \right\}$.
However, the registry index does not describe the \inter-GNR transmission in a general fashion.
To this extent we presented a simple model based on a description of the electronic states as plane waves that captures the effect of the angle. Furthermore, we show how our model explains the role of the intersection angle in crossed GNRs with different orientations.

The emerging picture from the combination of Ref.~\cite{Lima2016} and the work presented here, is that GNRs with different chiralities and widths may be combined in nanoscale crossbar junctions which should allow, under suitable control of the intersection angle, to construct effective and tunable electronic beam splitters.

\section{Acknowledgments}

The authors acknowledge financial support from FP7 FET-ICT ``Planar Atomic and Molecular Scale devices'' (PAMS) project (funded by the European Commission under contract No.~610446), the Spanish Ministerio de Economia y Competitividad (MINECO) (Grant No.~MAT2013-46593-C6-2-P), the Basque Dep.~de Educaci\'on and the UPV/EHU (Grant No.~IT-756-13).


\bibliographystyle{apsrev-title}
\bibliography{references}

\end{document}